\documentclass[
    ,final            
  ]
  {aipproc}

\layoutstyle{6x9}

\newcommand{\be}{\begin{equation}}
\newcommand{\ee}{\end{equation}}
\newcommand{ \bea }{\begin{eqnarray}}
\newcommand{ \eea }{\end{eqnarray}}
\newcommand{ \la }{\langle}
\newcommand{ \ra }{\rangle}

\begin{document}
\title{Effect of Rescattering on Forward-Backward Correlations}

\classification{24.60.Ky, 25.75.-q, 25.75.Dw, 25.75.Gz, 25.75.Nq, 12.38.Mh}
\keywords      {Heavy Ion Collisions, Event by Event Fluctuations, Color Glass Condensate, Long Range Correlations.}

 \author{Stephane Haussler}{
   address={Frankfurt Institute for Advanced Studies (FIAS), J.W. Goethe Universit\"at, \\
 Max von Laue Stra\ss{}e 1, 60438 Frankfurt am Main, Germany}
 }
 \author{Mohamed Abdel-Aziz}{
 address={Institut f\"ur Theoretische Physik, J.W. Goethe Universit\"at, \\
 Max von Laue Stra\ss{}e 1, 60438 Frankfurt am Main, Germany}
 }
 
 \author{Marcus Bleicher}{
 address={Institut f\"ur Theoretische Physik, J.W. Goethe Universit\"at, \\
 Max von Laue Stra\ss{}e 1, 60438 Frankfurt am Main, Germany}
 }
 
\begin{abstract}
We use the Ultra-relativistic Quantum Molecular Dynamic model (UrQMD
v2.2) to study forward-backward fluctuations and compare our
results with the data published by the PHOBOS experiment.
The extracted effective cluster multiplicities show a clear centrality
dependence within the present hadron-string transport approach.
This behavior is not reproduced with models not taking into
account final state rescattering.

\end{abstract}
\maketitle
\section{Introduction}\label{sec:introduction}

One of the main goal of the relativistic heavy ion program is to
understand the nature of hadron production mechanisms (e.g.\
parton coalescence, string fragmentation or cluster decay).
Numerous data suggest the formation of  a quark gluon plasma (QGP)
during the collision of two heavy nuclei at ultra-relativistic
energies. Using correlations and fluctuations to probe the nature
and properties of the highly heated, high density matter created
in the course of these collisions has been proposed by many
authors, see for
example \cite{Haussler:2005ei,Abdel-Aziz:2006fe,Haussler:2006rg,Li:2006eh,Gavin:2006xd,Jeon:2005kj,Koch:2001zn,Bleicher:2000ek,Bleicher:1998wu,Armesto:2006bv}.
In particular, one can get insight about cluster decay with the
help of such fluctuation studies.

The UA5 experiment performed a study of cluster size in $p-\bar{p}$ collisions
by analyzing forward-backward charged particle multiplicity fluctuations \cite{Alpgard:1983xp}.
They found a cluster multiplicity around two,
in line with the expected result in case clusters correspond to decaying resonances
(e.g $\rho^{0} \to \pi^{+}+\pi^{-}$).
A similar analysis was recently performed by the PHOBOS experiment
for $Au+Au$ reactions at $\sqrt{s_{NN}}=200$~GeV \cite{Back:2006id}.

In Refs. \cite{Abdel-Aziz:2006fe,Haussler:2006rg}, a simple model was introduced to extract
the effective cluster multiplicity $K_{\rm eff}$ from the PHOBOS data.
Within this approach, $K_{\rm eff}$ is found to be $2.7$ for mid-peripheral events
and $K_{\rm eff}\sim 2.2$ for central events.
The value of $K_{\rm eff}$ in central collisions is close to the $p-\bar{p}$ value reported by
UA5 \cite{Alpgard:1983xp}. Note that all measured cluster
multiplicities $K_{\rm eff}$ are larger than the one
computed for a hadron resonance gas ($K_{\rm HG}=1.5$)
\cite{Stephanov:1999zu}, indicating that the measured charge
correlations can not be described by simple statistical models
based on hadronic degrees of freedom. The STAR collaboration
has measured the long range correlations and it was shown that also
within a string fusions approach, the data for central $A+A$ reactions can not be
reproduced \cite{Tarnowsky:2006nh}.

In this study we calculate a baseline estimate for
forward-backward fluctuations based on the microscopic hadronic
transport model UrQMD. For a complete review of the model see
\cite{Bass:1998ca}. The centrality, rapidity $\eta$ and rapidity window $\Delta\eta$ dependence of  the dynamical fluctuations are studied and interpreted in term of effective cluster multiplicities.
For this analysis, $5\times10^5$ $pp$ and minimum bias Au+Au events at $\sqrt{s_{NN}}=200$~GeV
were used.

\section{Forward-Backward Fluctuations}\label{sec:model}
In this section, we introduce the variable $C$ that measures the
asymmetry between the forward and backward charges. We define two
symmetric rapidity regions at $\pm\eta$ with equal width
$\Delta\eta$. The number of charged particles in the forward
rapidity interval $\eta\pm\Delta\eta/2$ is $N_F$ while the
corresponding number in the backward hemisphere
$-\eta\pm\Delta\eta/2$ is given by $N_B$. We define the asymmetry
variable $C=(N_F-N_B)/\sqrt{N_F+N_B}$,
in each event. The variance of the charged particle multiplicity
in the forward hemisphere is given by $D_{FF}=\la N_F^2\ra-\la
N_F\ra^2$ and similarly for the backward hemisphere $D_{BB}=\la
N_B^2\ra-\la N_B\ra^2$. We also introduce the covariance of
charged particles in both hemispheres by $D_{FB}=\la N_FN_B\ra-\la
N_F\ra\la N_B\ra$, where $\la....\ra$ stands for the average over
all events. The PHOBOS measure of the dynamical fluctuations
$\sigma^2_C$ can be written as
\be \sigma^2_C=\la C^2\ra-\la C\ra^2\approx
\frac{D_{FF}+D_{BB}-2D_{FB}}{\la N_F+N_B\ra}.\ee

Recently, STAR \cite{Tarnowsky:2006nh} reported preliminary
results of the so called correlation strength parameter
$b=D_{FB}/D_{FF}$. The effective cluster multiplicity $K_{\rm
eff}$ is proportional to $\sigma^2_C$, such that if b=0, then the
covariance $D_{FB}$ vanishes. In this case we have
$\sigma^2_C=K_{\rm eff}$.
We emphasize that $K_{\rm eff}$ should be understood as a product
of the true cluster multiplicity times a leakage factor $\xi$ that
takes into account the limited observation window $\Delta\eta$ \cite{Abdel-Aziz:2006fe,Haussler:2006rg}.
The event by event fluctuations of the asymmetry parameter
(variance) $\sigma^2_C$ in the absence of any correlations among
the produced particles will be $\sigma^2_C=1$. If there is only
long range correlation, then $0<\sigma^2_C<1$.
\section{Comparison to data}
\begin{figure}
 \vspace*{-0.0cm}
\centerline{
\includegraphics[height=.25\textheight]{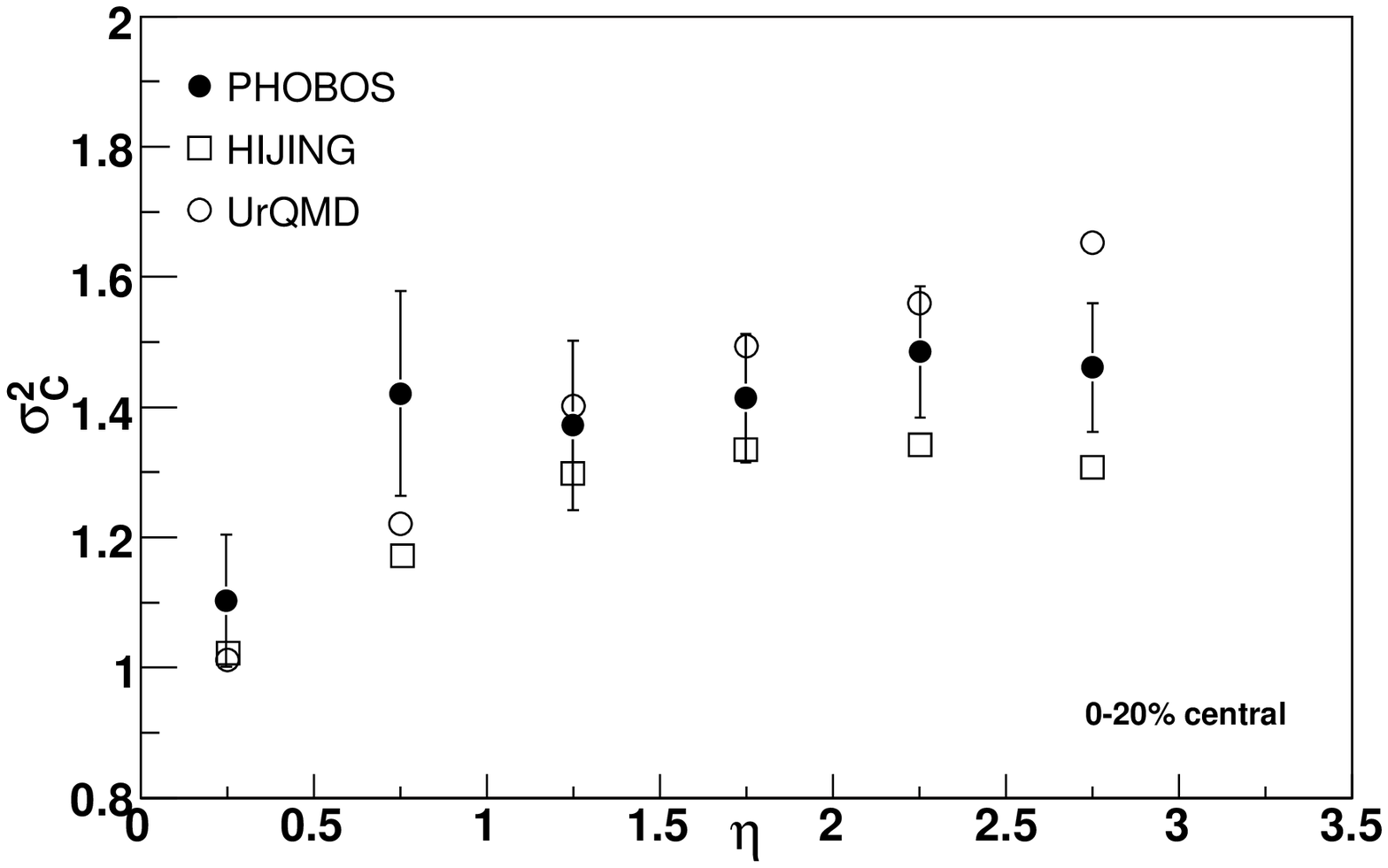}
    \hspace*{0.cm}
    \includegraphics[height=.25\textheight]{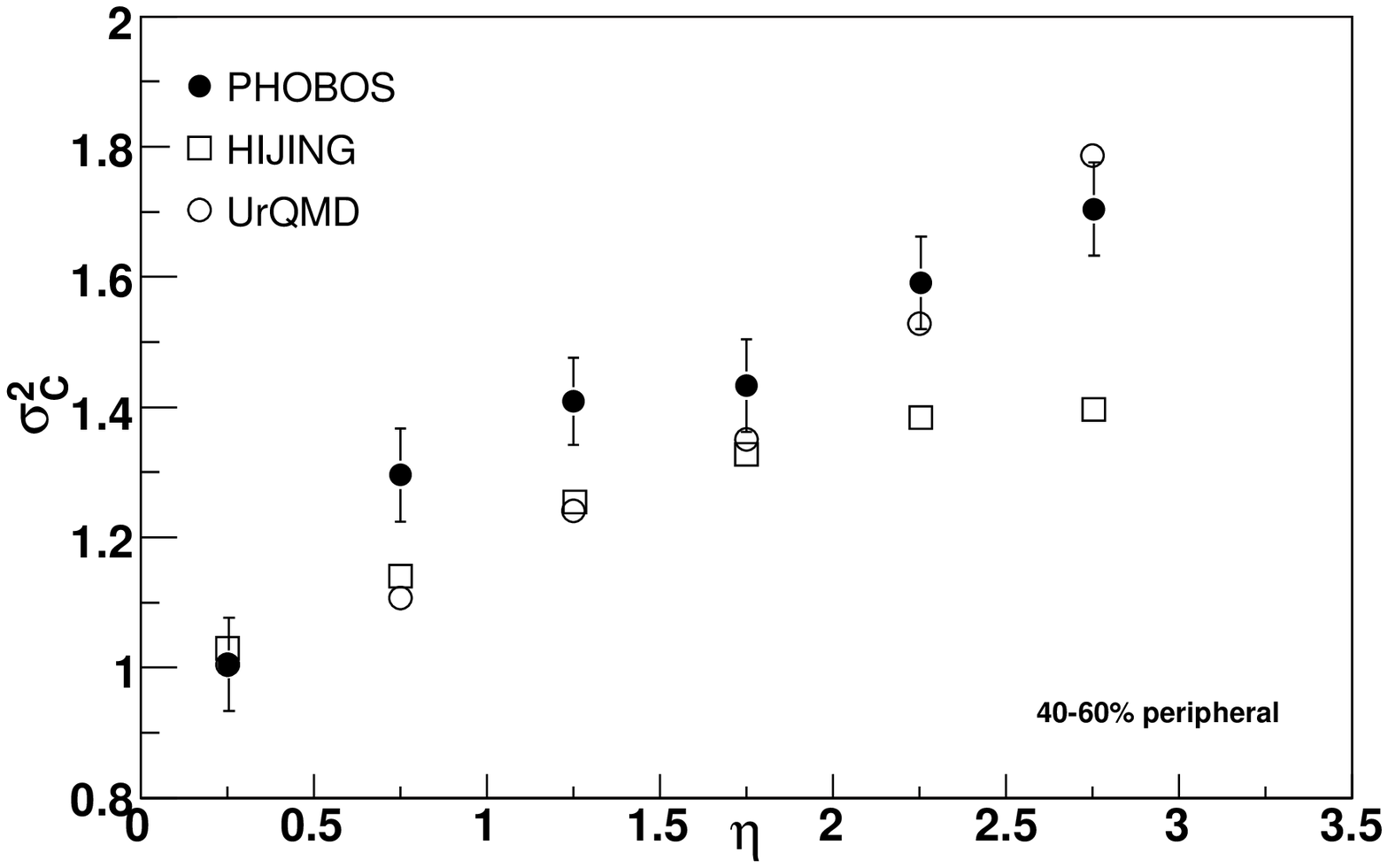}
} \vspace*{-0.95cm} \caption{Dynamic fluctuations as a function of
pseudo rapidity $\eta$ at $\sqrt{s_{NN}}=200$
 GeV, with $\Delta\eta=0.5$.
 (Left) $0\%-20\%$ central Au+Au, (right) 40\%-60 peripheral Au+Au collisions.
 Full circles are PHOBOS \cite{Back:2006id} data, open squares HIJING and open circles are UrQMD results.}
 \label{fig:1}
  \vspace*{-0.4cm}
\end{figure}

We compare the UrQMD model calculations with the existing experimental
data reported by PHOBOS for $Au+Au$ collisions at $\sqrt{s_{NN}}=200$~GeV in Figs.~\ref{fig:1},~\ref{fig:2}. Also shown here are the HIJING result as reported by PHOBOS \cite{Back:2006id}.

Fig.~\ref{fig:1} depicts $\sigma^2_C$ as a function of the pseudo-rapidity $\eta$
while the window is fixed at $\Delta \eta =0.5$. In Fig.~\ref{fig:2}, we show $\sigma^2_C$
as a function of $\Delta \eta$ while $\eta$ is fixed at $\eta = 2$.
Central data correspond to the $0\%-20\%$
most central events while peripheral data correspond to $40\%-60\%$ events as determined
from the number of charged particles at mid-rapidity.
We find that the present transport approach is able to reproduce mid-peripheral PHOBOS data roughly for both
pseudo-rapidity $\eta$ and window $\Delta \eta$ dependence (Fig.~\ref{fig:1}, right and Fig.~\ref{fig:2}, right). As a function of the pseudo-rapidity window $\Delta \eta$, UrQMD can mimic the mid-peripheral experimental data, however, it fails to reproduce the central one.

As a function of $\eta$ (Fig.~\ref{fig:1}),
$\sigma^2_C$ increases from $\sigma^2_C\approx 1$ to
$\sigma^2_C\approx 1.75$ for mid-peripheral events.
For central events, it increases from $\sigma^2_C\approx 1$ to $\sigma^2_C\approx 1.6$.
The value $\sigma^2_C\approx 1$ when $\eta=0.25$ can be explained by the competition between
long range and short range correlations which almost cancel out when the forward and backward
acceptances are very close. The negative long range component then decreases with $\eta$ and let $\sigma_C^2$ increase.

Fig.~\ref{fig:2} depicts the dependence of $\sigma^2_C$ on the size of the rapidity
window $\Delta \eta$.
The experimental value of $\sigma_C^2$ increases up to 2.8 for mid-peripheral events
and reaches $\sigma_C^2=2.2$ for central events.
UrQMD result overshoots PHOBOS data for central events.
In Ref. \cite{Abdel-Aziz:2006fe,Haussler:2006rg}, we argue that by increasing the
observation window $\Delta\eta$ one can see the whole cluster structure.
The only process able to destroy the cluster structure in UrQMD is hadronic rescattering.
The failure to reproduce central data can thus be seen as an indication for a lack of rescattering in UrQMD.

HIJING does not yield any centrality dependence and does not correctly reproduces the data for peripheral events,
however, surprisingly reproduces the central data.


\section{centrality dependence }

\begin{figure}
 \vspace*{-0.0cm}
\centerline{
    \includegraphics[height=.25\textheight]{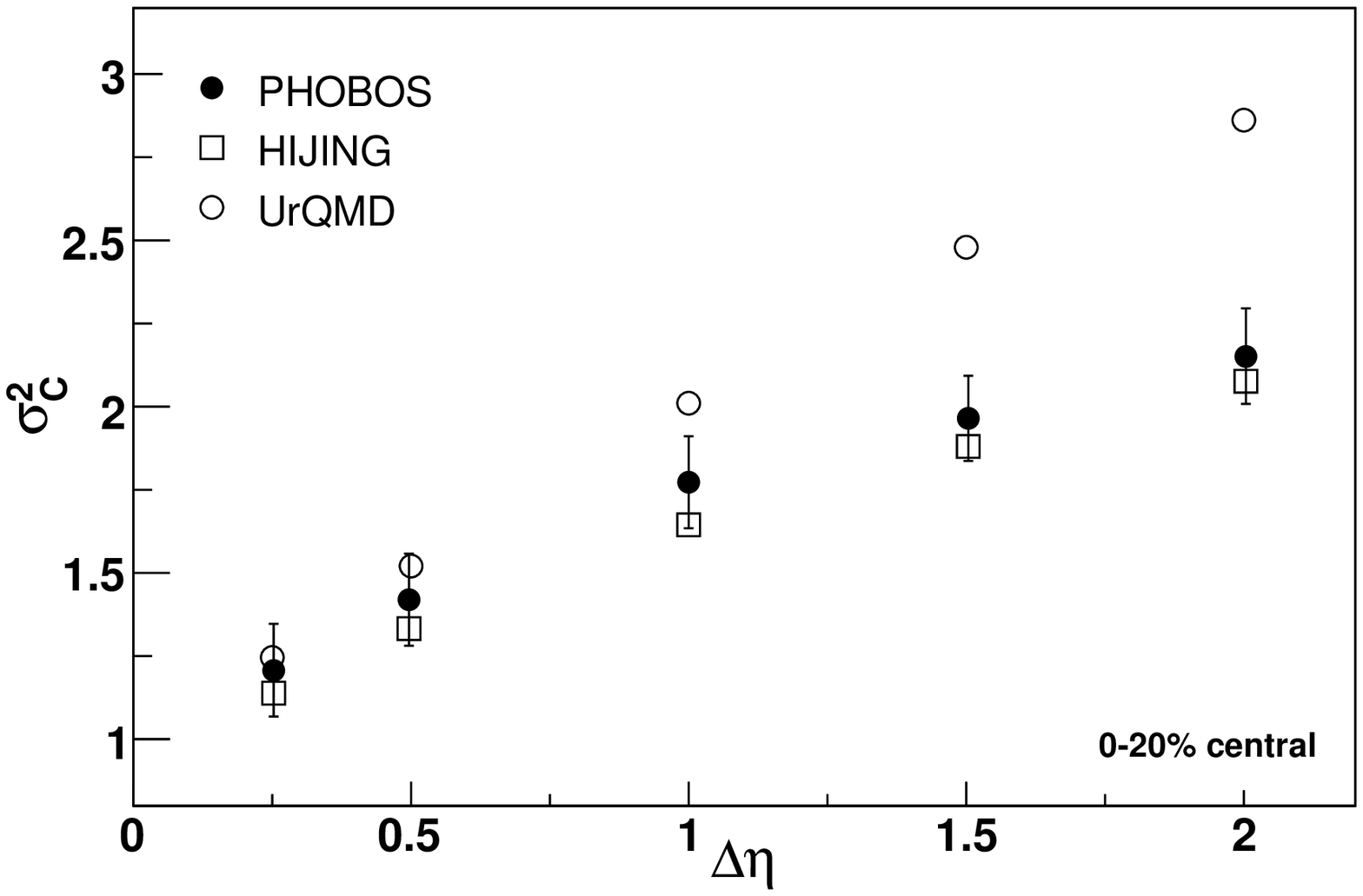}
    \hspace*{0.cm}
    \includegraphics[height=.25\textheight]{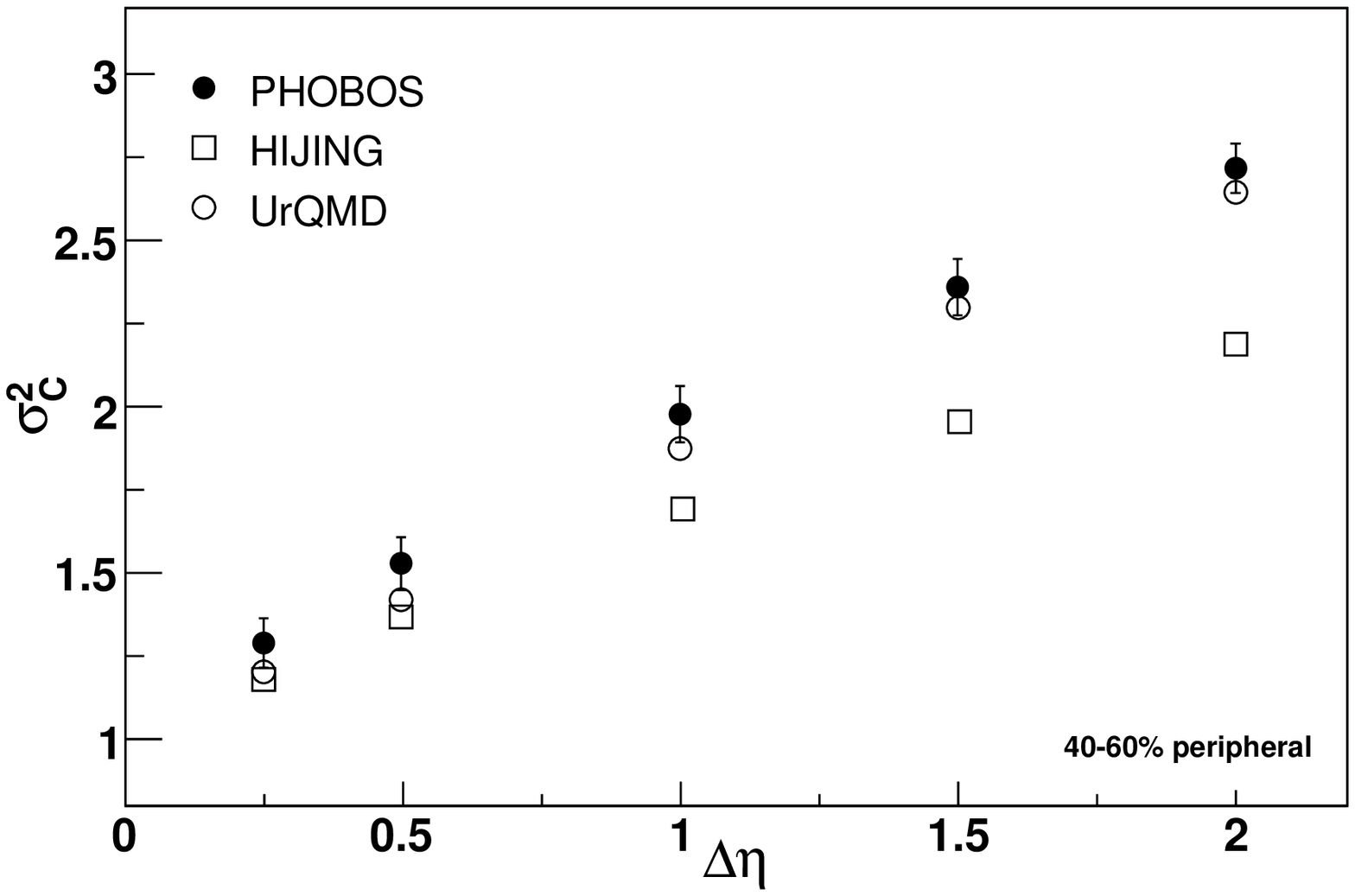}
} \vspace*{-0.95cm}
 \caption{Dynamic fluctuations as a function of
pseudo rapidity window $\Delta\eta$ at $\sqrt{s_{NN}}=200$
 GeV, with $\eta=2$
 (Left) $0\%-20\%$ central Au+Au, (right) 40\%-60 peripheral Au+Au collisions.
 Full circles are PHOBOS data \cite{Back:2006id}, open squares HIJING and open circles are UrQMD
 results.}
 \label{fig:2}
  \vspace*{-0.4cm}
\end{figure}

\begin{figure}
 \vspace*{-0.0cm}
\centerline{
    \includegraphics[height=.25\textheight]{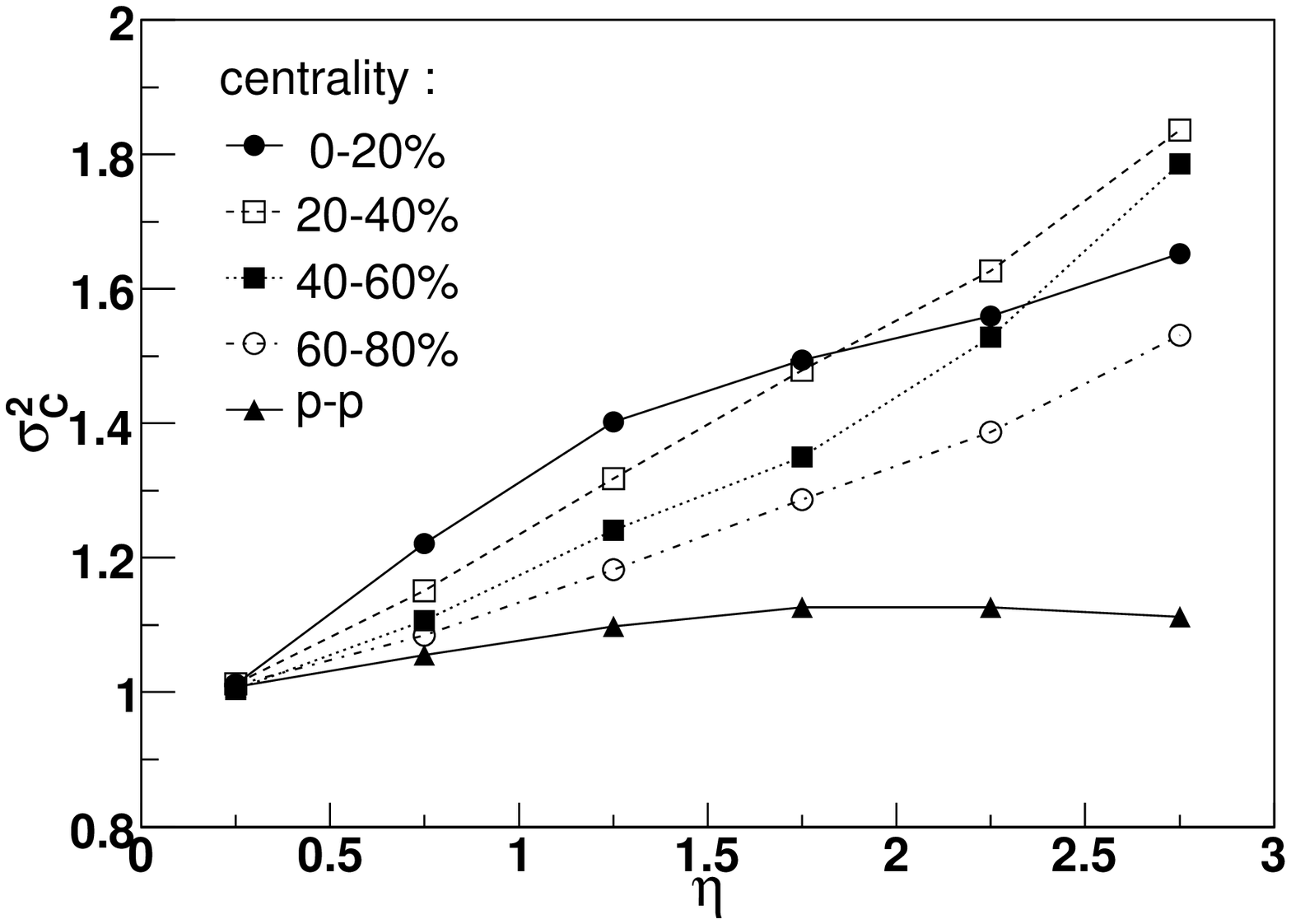}
    \hspace*{0.cm}
    \includegraphics[height=.25\textheight]{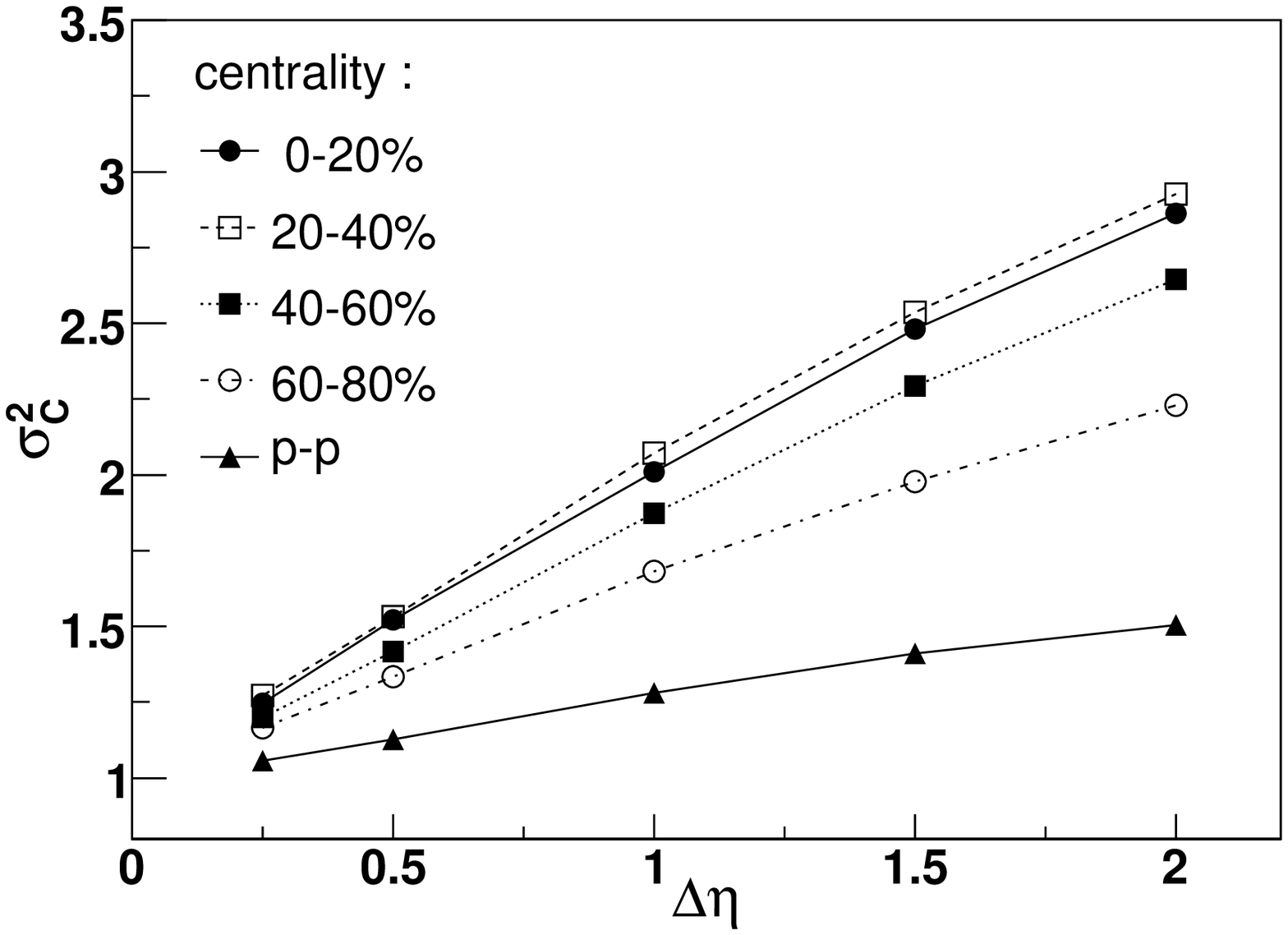}
} \vspace*{-0.95cm}
 \caption{UrQMD results of dynamic fluctuations as a function of pseudo rapidity $\eta$ (left, $\Delta \eta=2$) and the observation window $\Delta \eta$ (right, $\eta=2$).}
 \label{fig:3}
 \vspace*{-0.4cm}
\end{figure}

In this section, we study the centrality dependence of the forward-backward fluctuations calculated from UrQMD for a set
of $Au+Au$ and $p+p$ events at the highest RHIC energy available.
$\sigma^2_C$ as a function of $\eta$ is shown in Fig.~\ref{fig:3} (left) for different centralities. We observe a clear increase of $\sigma^2_C$ from $p+p$ ($\sigma^2_C=1.1$ at $\eta=2.75$) up to the $20-40\%$ $Au+Au$ most central events ($\sigma^2_C=1.8$ at $\eta=2.75$).
With even higher centrality, the behavior then changes and $\sigma^2_C$ gets now smaller ($\sigma^2_C=1.6$ at $\eta=2.75$ for the $0-20\%$ most central events). The same trend is observed as a function of the pseudo-rapidity window $\Delta \eta$
shown in Fig.~\ref{fig:3} (right).
$\sigma^2_C$ increases from $p+p$ ($\sigma^2_C=1.25$ at $\Delta \eta=2$) to the $20-40\%$ $Au+Au$
most central events ($\sigma^2_C=2.75$ at $\Delta \eta=2$).
In \cite{Abdel-Aziz:2006fe,Haussler:2006rg} we predicted that for PHOBOS data, $\sigma^2_C$ may be reduce to 1.9
with tighter centrality cuts.
This reduction in $\sigma^2_C\sim K_{\rm eff}$ may be regarded as an indication
for cluster melting at RHIC.

\begin{figure}\label{fig:4}
\vspace*{-0.0cm} \centerline{
    \includegraphics[height=.25\textheight]{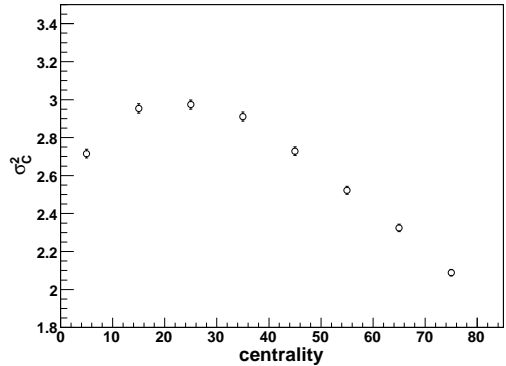}
    \hspace*{0.cm}
}\vspace*{-0.95cm}
\caption{ Dynamical fluctuations computed with UrQMD as a function of centrality , with $\eta=2$,
 and $\Delta\eta=2$.
}
\vspace*{-0.4cm}
\end{figure}

A summary of the centrality dependence is presented in Fig.~\ref{fig:4}, where $\sigma^2_C$ is shown for fixed $\eta=2$ and $\Delta \eta=2$.
With this acceptance window, $\sigma^2_C$ increases from $\sigma^2_C=2.1$ for peripheral events up to $\sigma^2_C=3$ for the $20-30\%$ most central events. Followed by a decrease down to $\sigma^2_C=2.7$ for the $0-10\%$ most central events.
From Fig.~\ref{fig:2}, we see that a model without final state rescattering like HIJING
seem not to be able to mimic the decreasing cluster size with the most central events.

\section{Summary and Conclusion}\label{sec:summary}

In this paper we computed forward-backward fluctuations and
compared UrQMD calculations results to the available experimental
data measured by the PHOBOS collaboration. The study of proton-proton
collisions indicated that the long range correlation persists
over a wide rapidity gap between the two rapidity hemispheres. The
variance of the asymmetry parameter $C$ was found to increase with
increasing $\Delta\eta$ such that $\sigma^2_C$ changes from
$\sigma^2_C(\eta=2,\Delta\eta=0.25)\approx 1$ to
$\sigma^2_C(\eta=2,\Delta\eta=3)\approx 1.6$, this can be due the
saturation of the leakage factor $\xi\rightarrow 1$.

For Au+Au collisions, we found that for both centrality bins
$0\%-20\%$ and $40\%-60\%$, $\sigma^2_C\approx 1$ for small
$\eta$. This can be seen as a cancellation between short and long
range correlations. By increasing $\eta$ we observed that $\sigma^2_C$
also increases and approaches 1.6 and 1.8 for $0\%-20\%$ and
$40\%-60\%$, respectively. This increase can be attributed to the
decrease in the long range correlations. This will be true if the
particle production mechanism does not change with $\eta$. To see
the whole cluster structure, we fixed the center of the observation
window at 2 and allowed $\Delta\eta$ to increase. We found that UrQMD
can reproduce the peripheral data while it overestimates the
experimental results for central collisions. This might indicate an additional cluster melting process
not accounted for in the present hadron-string model.

\begin{theacknowledgments}
This work is supported by BMBF and GSI. Computational resources have been provided by the Center
of Scientific Computing (CSC) at J.W. Goethe University.
\end{theacknowledgments}

\end{document}